# 1 Magnetic rotational spectroscopy for probing rheology of nanoliter droplets and thin films

Konstantin G. Kornev, Yu Gu, Pavel Aprelev, and Alexander Tokarev

Department of Materials Science and Engineering, Clemson University, SC

# 2 Magnetic rotational spectroscopy

# 3 Definition of the topic

As one of the characterization tools for nanoscience & nanotechnology, magnetic rotational spectroscopy (MRS) with magnetic nanoprobes is a powerful method for *in-situ* characterization of minute amounts of complex fluids and for study of nanoscale structures in fluids. In MRS, a uniformly rotating magnetic field rotates magnetic micro- or nano-probes in the liquid and one studies rheological properties by analyzing the features of the probe rotation. The technique relies only on full rotations (rather than oscillations) of the magnetic moment of the probes.

# 4 Abstract

In-situ characterization of minute amounts of complex fluids is a challenge. Magnetic Rotational Spectroscopy (MRS) with submicron probes offers flexibility and accuracy providing desired spatial and temporal resolution in characterization of nanoliter droplets and thin films when other methods fall short. MRS analyzes distinct features of the in-plane rotation of a magnetic probe, when its magnetic moment makes full revolution following an external rotating magnetic field. The probe demonstrates a distinguishable movement which changes from rotation to tumbling to trembling as the frequency of rotation of the driving magnetic field changes. In practice, MRS has been used in analysis of gelation of thin polymer films, ceramic precursors, and nanoliter droplets of insect biofluids. MRS is a young field, but it has many potential applications requiring rheological characterization of scarcely available, chemically reacting complex fluids.

# 5 INTRODUCTION

Recent focus in the science of soft materials has moved to the analysis of mechanisms that drive the assembly of small building blocks such as polymers, nanoparticles, and nanofibers into an enormous diversity of multifunctional materials. The properties and performance of these materials significantly depend on the material organization at different scales. It has been shown that the inter-scale interactions are manifested through the response of the materials to the applied loads. It appears that the rheological parameters of materials, such as viscosity, relaxation times, and elastic moduli, carry all the necessary information about *in vivo* physico-chemical reactions and mechanisms of polymer, colloid, and nanofiber assembly and gelation in the real time [1-15]. Minute samples are most attractive for tracking inter-scale interactions and "cross talk" between scales. These "cross talks" are especially important in understanding and analysis of self-assembly of biomaterials into small organisms that range from cells to embryos to insects [2-4, 15, 16]. Characterization of rheological properties of small samples presents a significant challenge that requires special methods and techniques. Therefore, microrheology is distinguished from the main stream of rheology research and constitutes a separate field of materials science.

When the material rapidly changes its rheological properties due to changes in the environmental conditions, its characterization becomes especially challenging [2, 5]. For example, in biofluids, such as cellular, mucosal and tissue fluids, the fluid viscosity changes over time making it challenging to quantify and analyze these samples. Rheology of polymer solutions and gels depends on the concentration, level of cross-linking, oxygen content, temperature, pH, and many other environmental parameters. Drastic thickening of fluids in a short time interval often results in the far-reaching consequences. One example of this is sickle cell anemia, when the cytosolic viscosity inside the red blood cells changes drastically [17-19]. Another is polymerization of fibrinogen during fibrin clot formation and wound healing[20-22]. Furthermore, many polysaccharides gel in fractions of a second; this phenomenon is actively used by animals and insects to make reactive extracts and cuticular building blocks[3, 4, 23]. Recently, these polysaccharides found practical applications in manufacturing fibers and films [24-28] and attract attention of microrheologists [29-34].

To address the challenges of *in situ* characterization of rheological properties of materials, different experimental methods have been proposed and developed. In many cases, rheological characteristics of materials are inferred by comparing the translational and/or rotational motions of different tracers against available models of particle-medium interactions [29, 30].

The tracer - or probe - is considered passive when its motion is caused by the forces exerted by the surrounding medium. The probe is considered active when it actively deforms the medium and is used to transfer the load onto the medium[31-33]. Small passive tracers are subject to thermal excitations and hence randomly move through the surrounding material. The mean squared displacement (MSD) of probes can be directly measured using the light-scattering techniques [14, 31, 35]. The rheological properties of the material are therefore extracted from the MSD by using a model of Brownian motion [1, 2, 35]. For example, applying the Stokes-Einstein relation for a spherical tracer moving in a Newtonian fluids, one can infer fluid viscosity.

Many materials can be made magnetic by dispersing magnetic micro and nanoparicles in them. These micro and nanoparticles are available on market or can be produced in the laboratory (see, for review, [36-43]). Magnetic particles can be considered active probes because they can be put in motion by applying an external magnetic field. The idea of using magnetic tracers to probe rheological properties of materials was originated from the pioneering work of Crick and Hughes[44, 45]. Crick and Hughes used magnetic particles to probe viscosity and elastic reaction of cytoplasm. An applied rotating magnetic field exerts a torque on a magnetic particle, which is balanced by the viscous and elastic torques acting from the medium. Crick and Hughes studied the reaction of the medium on a step-like pulse of the external magnetic field. They monitored the particle relaxation to its equilibrium position. Following their ideas, magnetic tracers, mostly spherical micro- and nano-particles, have been used in different applications[46-57].

Magnetic tracers are typically tracked by either directly filming the tracer movement or using some indirect methods of the particle detection. Examples of indirect methods include measurements of AC susceptibility[55-57], coercivity[58], remanence[46, 59, 60], or light intensity [61-63]. The rheological properties of materials are obtained by comparing experimental data with predictions of appropriate models of particle/medium interactions [29, 47-52, 64-72]. Anisotropic particles such as wires, rods, and chains have attracted attention of microrheologists only recently[65, 67-71, 73-80]. Magnetic nanorods have several advantages over spherical nanoparticles: due to their anisotropic shape, rotational motions of nanorods can be easily tracked and analyzed from the microscope images. Moreover, magnetization of a rod-like particle is often codirected with the rod axis [81, 82]. This fact significantly simplifies the models of nanorod rotation making rheological measurements reliable.

Magnetic rotational spectroscopy (MRS) takes advantage of a distinguishable behavior of rotating tracers as the frequency of applied rotating field changes. Unlike many methods based on the analysis of small oscillations, MRS with magnetic nanorods enjoys analysis of full revolutions of magnetic tracers, which is much easier to track using inexpensive microscopes. Nanorods can be kept strictly in the focal plane of the microscope by controlling the applied magnetic field. Moreover, nanorods as thin as hundreds of nanometers in diameter can be seen with dark field imaging. Therefore, MRS with magnetic nanorods provides very accurate data on submicron rheology of materials [67-69, 71, 83].

As first shown by Frenkel [84], rotation of a rod-like particle in a Newtonian fluid changes from synchronous, when the rod continuously follows the rotating field, to asynchronous, when the rod periodically swings back and forth. This transition occurs at a certain frequency of the rotating magnetic field. Frenkel's effect was actively employed in the last century to study rod-like polymers and liquid crystals[1, 2]. Direct observations of the critical behavior of rod-like particles, however, were lacking, and all measurements were conducted indirectly. The critical transition from synchronous to asynchronous rotation was first visualized only in 2005 when carbon nanotubes filled with magnetic nanoparticles were used for these purposes[85]. Transition from synchronous to asynchronous rotation of magnetic nanotubes was used to estimate magnetic properties of composite nanotubes.

These experiments pushed forward the development of Magnetic Rotational Spectroscopy as a new method to study rheological properties of nanoliter droplets [50, 51, 64, 68, 69, 71, 74, 76, 83, 86-88]. The idea of using the transition from synchronous to asynchronous rotation was implemented by several groups employing different magnetic probes.

In this review, we will focus on MRS with the rod-shaped particles. Rheological characterization of simple and complex fluids will be discussed. We will review the current state of the art in this field and outline its future perspectives.

## 5.1 Magnetic rotational spectroscopy for Newtonian fluids

The MRS theory for simple fluids has been first developed by Frenkel[84] who noticed that a magnetic rod synchronously rotates with the applied rotating magnetic field only within a definite window of the applied frequency. This effect can be explained by analyzing a simple model of rotation of a single magnetic particle in a 2D rotating magnetic field. In Newtonian fluids, the viscous drag is linearly proportional to the particle angular velocity[89], provided that the Reynolds number Re = $\rho R^2 f/\eta$ is very small, where $\rho$ is the fluid density, R is the particle size, f is the rotation frequency of applied magnetic field, and $\eta$ is the fluid viscosity. In MRS applications, this is always the case and can be confirmed with a simple estimation. Taking as an upper estimate for the particle size R = 10µm and for the rotation frequency f = 10 Hz, we have for water, Re = $\rho R^2 f/\eta$ = $10^3 \cdot 10^{-10} \cdot 10 / 10^{-3}$ = $10^{-3}$. More viscous fluids will provide even smaller Reynolds numbers. Thus, a linear relation between the viscous torque and rotation rate of the particle should be expected.

## 5.2 FERROMAGNETIC PARTICLES

Consider first the case of ferromagnetic particles. We assume that the magnetic moment **m** is fixed at an easy axis of the ferromagnetic particle. The angle $\alpha$ specifies the direction of the applied rotating magnetic field **B** with respect to the reference axis X. This angle depends linearly on time, $\alpha = 2\pi f t$, where $f$ is the frequency of the rotating magnetic field. The drag force resists the particle rotation causing the magnetic moment **m** to lag behind the field **B** with angle $\theta$ (Fig. 1). In order to derive the equation governing a particle rotation, it is convenient to count its revolutions with respect to the fixed system of coordinates, i.e. with respect to the X-axis. Therefore, if the easy axis of a particle makes angle $\varphi(t)$ with the X-axis, the torque balance equation reads [90-93]

$$(\gamma d\varphi/dt)\mathbf{e} = \mathbf{m} \times \mathbf{B}, \tag{1}$$

where $\gamma$ is the drag coefficient and **e** is the unit vector directed perpendicularly to the plane of the particle rotation. Substituting the definition of angle $\varphi(t)$ through the angles $\alpha(t)$, and $\theta(t)$, $\varphi(t) = 2\pi f t - \theta(t)$. The governing equation thus takes on the following form:

$$\gamma\left(2\pi f - \frac{d\theta}{dt}\right) = mB\sin\theta, \tag{2}$$

where the drag coefficients for different particles are defined as [1]

$$\gamma = 8\pi R^3 \eta \text{ (sphere of radius R)}, \tag{3}$$

$$\gamma = \frac{\pi l^3 \eta}{3\ln(l/d) - A} \quad \text{(rod of length } l \text{ and diameter } d\text{), A = 2.4.} \tag{4}$$

As illustrated in Fig. 1 e), equation (1) has a steady state solution where d$\theta(t)$/dt =0. This implies that the particle rotates with the frequency of the applied rotating field, but the direction of magnetization vector does not necessarily coincide with the field direction making a finite angle θ with the field.

However, if the frequency becomes greater than a certain critical frequency given by equation $2\pi f_c = \omega_c = mB/\gamma$, this solution disappears and the particle cannot rotate in unison with the field anymore[84]. One observes that the particle oscillates and the angle θ formed by the magnetization vector and the field vector changes with time. Fig. 1f) illustrates this behavior: periodic solutions correspond to asynchronous rotation of magnetic nanorods.

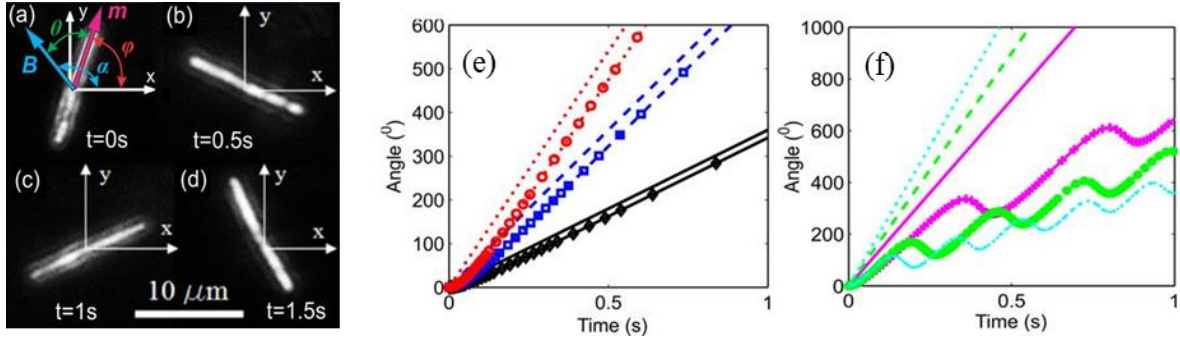

Figure 1. Images of the nickel nanorod (200nm in diameter) rotating anti-clockwise. The 2D uniform magnetic field rotates with constant frequency f=0.5Hz forcing the nanorod to follow the field [69]. (e) The solutions of Eq. 1 for different driving frequencies of magnetic field. The lines without any symbols correspond to the change of $\alpha$ - angles with frequencies f = 1Hz (straight solid line), f = 2Hz (straight dashed line) and f = 3Hz (straight dotted line). The corresponding φ-solutions are marked with different symbols. The nanorod rotates synchronously with magnetic field at these frequencies, (f) illustration of asynchronous rotation showing drastic difference in behavior of the solutions to Eq. 1 above the critical frequency f = 3Hz. The straight lines correspond to the change of the alpha-angles with frequencies f=4Hz (straight solid line), f=5Hz (straight dashed line) and f=6Hz (straight dotted line). The corresponding φ-solutions are shown in the same color and the lines are not straight as opposed to those in (a). Parameters of numerical experiment: nanorod length l=6.6 µm, fluid viscosity η = 16·10$^{-3}$Pa·s, magnetization M=2.25*10$^{-14}$ A*m$^2$ and magnetic field B=0.0015 T [67].

Expressing magnetic moment through the material magnetization M, $m_{sphere} = (4/3)\pi MR^3$, $m_{rod} = \pi M l d^2/4$, we notice that the dimensionless critical frequencies, $f_d$, introduced as

$$f_d = \frac{8\pi f_c \eta}{MB} = \frac{2}{3}, \quad \text{(sphere),} \tag{5}$$

$$f_d = \frac{8\pi\eta f_c}{MB} = \left(3\ln\left(\frac{l}{d}\right) - 2.4\right)\left(\frac{l}{d}\right)^{-2}, \text{ (rod)}, \tag{6}$$

do not depend on the particular size of the sphere and nanorod. For a spherical particle this critical frequency is constant, but for a rod-like particle, this frequency depends only on the length-to-diameter ratio $l/d$ of the rod [68, 69, 83, 94] (Fig. 2b). This effect of the rod-shaped particle is explained as follows: the viscous torque in eqs. (2) and (4) is proportional to $l^3$, while the magnetic torque is proportional to $d^2l$. These two torques compensate each other at the critical frequency, resulting in a universal dependence (6). This dependence plays an important role when one needs to choose a particular magnetic probe. It appears, that one can select either material magnetization ($M$) or the rod aspect ratio ($l/d$) and operate within a particular frequency band to enable measurements of viscosity of the wide range of fluids [69, 73, 83].

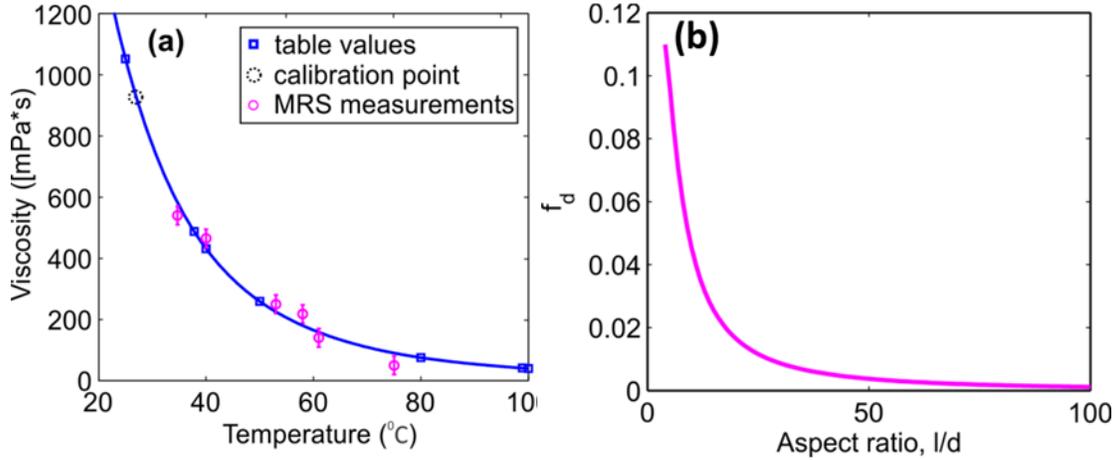

Figure 2. Viscosity of the liquid standard S600 measured by MRS with the nanorods with different aspect ratios; squares and blue line show the table values of viscosity; open circles show measured viscosity from 5 independent experiments with nanorods of different aspect ratios. b) Theoretical dependence of dimensionless critical frequency $f_d$ on the nanorod aspect ratio[67].

Our group confirmed this theory by using different fluids and different magnetic materials [69]. As an illustration of the excellent agreement of the theory with experiment, we present Fig. 2a) where a standard liquid with a temperature dependent viscosity was used in the MRS experiments.

## 5.3 PARAMAGNETIC RODS

Upon application of a rotating magnetic field on a paramagnetic sphere, magnetic moment rotates within the sphere to follow the changing field. Therefore, a spherical particle does not interact with its surrounding non-magnetic fluid. In contrast, magnetic moment of a high aspect ratio paramagnetic rod does not deviate significantly from the rod axis due to the difference in the axial and radial susceptibilities of the rod-like particles[76, 81]. Hence the rod-like particles do interact with the surrounding fluid. The basic equations (1) and (2) are re-written for the paramagnetic rods as [76]

$$\gamma \frac{d\varphi}{dt} = \frac{\pi d^2 l}{2\mu_0} \frac{\chi^2}{2+\chi} B^2 \sin(2\theta), \text{ or}$$

$$\frac{d\varphi}{dt} = \omega_c \sin(2\theta), \quad \omega_c = \frac{1}{\gamma} \frac{\pi d^2 l}{2\mu_0} \frac{\chi^2}{2+\chi} B^2 \quad \text{(Paramagnetic)},$$

(7)

where χ is the susceptibility of the rod material. In the rotating magnetic field, the critical frequency 2πf_c = ω_c separates synchronous (ω < ω_c) and asynchronous (ω > ω_c) rotations of paramagnetic rods. Experimentally, parameter ω_c is determined by observing the transition between these two distinct regimes of rotation.

## 5.4 TIME DEPENDENT VISCOSITY

When material's viscosity is time dependent, we cannot use eqs (5)-(7) anymore. One requires to take special care for modeling and studying the MRS features. Because the time dependent viscosity of materials is difficult to model, the majority of related studies are based on simple empirical dependencies [2, 95]. For example, many monomers undergoing polymerization is typically described by the following empirical equation [96, 97]':

$$\eta(t) = \eta_0 e^{\frac{t}{\tau}},$$

(8)

where $\eta_o$ is the initial viscosity of monomer solution, $t$ is the time and $\tau$ is the characteristic time of polymerization. This characteristic time is considered as a phenomenological parameter of the polymerization process [69, 97]. It is convenient to analyze this case by introducing the dimensionless times $T = 2\pi ft$ and $T_0 = 2\pi f\tau$. Since the form of the basic equation for the ferromagnetic and paramagnetic rods are very much similar to each other, we restrict ourselves to the analysis of rotation of ferromagnetic rods only. Equation (2) can be rewritten as: $\beta e^{T/T_0}(1-d\theta/dT) = \sin(\theta)$, where

$$\beta = \frac{\eta_0 l^3 2\pi f}{mB(3\ln(l/d) - A)}. \tag{9}$$

After making this nondimensionalization, all physical parameters collapse into two dimensionless complexes, $\beta$ and $T_0$. The dimensionless parameter $\beta$ describes all possible scenarios of the rod rotation.

Assume first that the drag force is much greater than the magnetic torque. This implies that the dimensionless parameter β is much greater than one, $\beta \gg 1$. Hence, as follows from the governing equation, $(1 - d\theta/dT) = (1/\beta)e^{-T/T_0}\sin(\theta) \approx 0$, or $\theta \approx T$, one would not expect to observe any rotation of the magnetic rod. In other limit, when the drag force is much smaller than the magnetic torque, the β-parameter is much smaller than one, $\beta \ll 1$. As follows from the governing equation, $\beta e^{T/T_0}(1 - d\theta/dT) = \sin(\theta)$, one would expect to observe the rod rotation until the left hand side is smaller than one.

Complete analysis of the rod rotation can be done using the phase diagram technique.[69] Introducing new auxiliary function $U = \beta e^{T/T_0}$, eq (2) is rewritten as a system of two first order differential equations,

$$\begin{aligned} d\theta/dT &= 1 - \sin\theta/U, \\ dU/dT &= U/T_0. \end{aligned} \tag{10}$$

The initial conditions satisfying the inequality $U_0 = \beta < 1$ correspond to the case when the magnetic torque is stronger than that caused by the viscous drag. In Fig. 3, the shaded region under curve $U = \sin\theta$ corresponds to the conditions that cause the rod to sway toward the field direction at the initial instants of time. As the time increases and the liquid gets thicker and thicker, the viscous drag takes over, and the rod slows down its rotation and eventually stops. Since magnetic field keeps rotating, the angle $\theta(U)$ increases with each revolution even if the rod is not moving. Therefore, the integral curves coming out from the shaded region describe the non-rotating rods. The shaded regions in Fig. 3 correspond to the initial orientations of the rods when they can be easy aligned with the field before the liquid thickens.

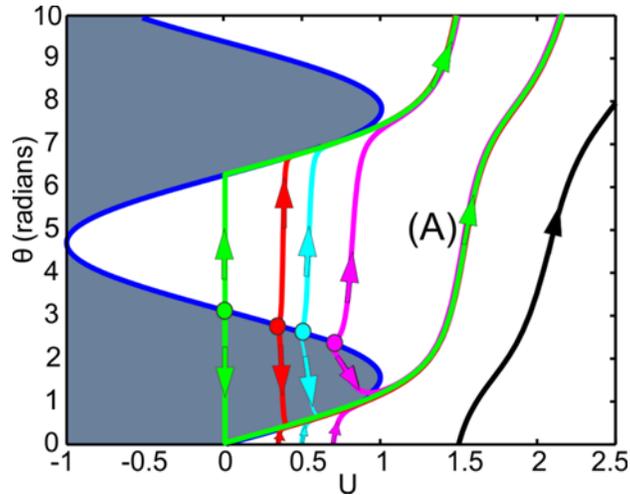

Figure 3. Phase portrait for eqs (10) showing the solution behavior for different initial conditions $U_0$, $\theta_0$ at fixed dimensionless $T_0 = \pi$. Only U > 0 is of practical interest hence the integral curves for U < 0 are not shown [69].

## 5.5 VISCOELASTICITY

Constitution of the cellular materials and rheological behavior of these materials challenged biophysicists and biologists for decades. Magnetic particles where employed by different groups to study viscoelastic behavior of biomaterials inside cells[46, 53, 98-101]. Rheological properties of cellular materials were extracted by studying the relaxation of the remanence field[46, 53, 98, 99] or by tracking the translational motion of a single particle[54, 102]. In MRS applications, where the rotational motion of a single particle is studied, one requires to determine the characteristic features of the particle rotation in a non-Newtonian viscoelastic fluid.

In this analysis, we will discuss only 2D rotation of magnetic rods in the viscoelastic Maxwell and Kelvin-Voigt fluids [2, 95]. In both models, a dash pot experiences viscous friction (viscosity: $\eta$) and a spring provides elastic reaction (elastic modulus: $G$) on the applied load.

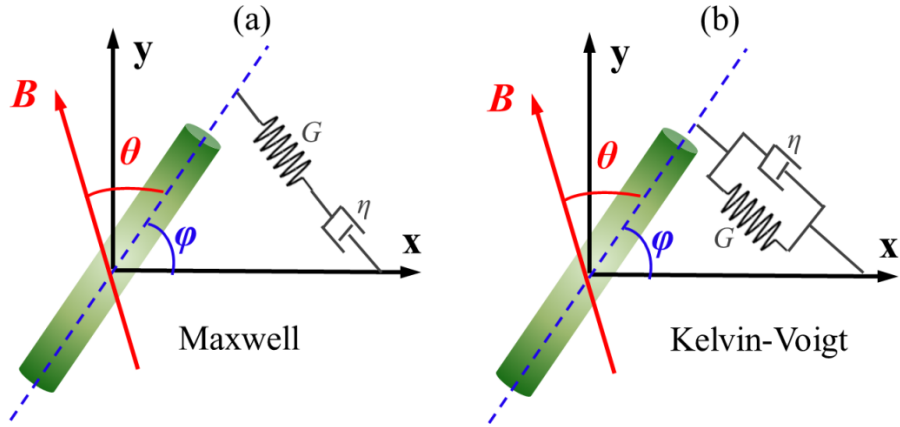

*Figure 4. Maxwell and Kelvin-Voigt models of the viscoleasticic reaction of a rod-like magnetic particle subject to magnetic field **B** (adapted from Ref. [71]).*

For the Maxwell model, the viscous $\tau_\eta = \gamma d\varphi_\eta/dt$ and elastic $\tau_G = \gamma G\, \varphi_G/\eta$ torques must be equal to the magnetic torque, $\tau_m = \pi d^2 l M B \sin\theta$ (ferromagnetic), or $\tau_m = \left(\pi d^2 l/(4\mu_0 + 2\mu_0 \chi)\right)\chi^2 B^2 \sin(2\theta)$ (paramagnetic): $\tau_\eta = \tau_G = \tau_m$. The angular displacement satisfies the relation: $\varphi_\eta + \varphi_G = \varphi$. Following these two relations, the equation governing the rotation of a rod-like particle reads [71]:

$$\frac{d\varphi}{dt} - \frac{\eta}{\gamma G}\frac{d\tau_m}{dt} = \frac{\tau_m}{\gamma}. \tag{11}$$

For the Kelvin-Voigt model, the torques satisfies the relation: $\tau_m = \tau_\eta + \tau_G$ and the angular displacement satisfies the relation: $\varphi_\eta = \varphi_G = \varphi$. The basic dynamic equation can be written as:

$$\frac{d\varphi}{dt} + \frac{G}{\eta}\varphi = \frac{\tau_m}{\gamma}. \tag{12}$$

We will limit ourselves to the discussion of behavior of ferromagnetic rods only. In this case, magnetic torque takes on the form $\tau_m = \pi d^2 l M B \sin\theta$ and the basic equations are written as follows[71]:

$$\frac{d\varphi}{dt} - \frac{\omega_c}{\omega_r}\cos\theta\frac{d\theta}{dt} = \omega_c \sin\theta \quad \text{(Maxwell Model)}, \tag{13}$$

$$\frac{d\varphi}{dt} + \omega_r \varphi = \omega_c \sin\theta \quad \text{(Kelvin-Voigt Model)}, \tag{14}$$

where $\omega_c = MBV/\gamma$ is the critical frequency (the same meaning as that of a Newtonian fluid) and $\omega_r = G/\eta$ is the reciprocal to the viscoelastic relaxation time. In the Maxwell model, the $\omega_r$ term accounts for the additive elastic resistance of the material to the rod rotation. As the viscoelastic relaxation time decreases (or $\omega_r$ increases), $\omega_r = G/\eta \rightarrow \infty$, the elastic resistance of the material to the rod rotation becomes much smaller than its viscous resistance. Consequently, angle $\varphi$ in Fig. 4 changes mostly due to the dash pot movement and eq.(13) reduces to eq.(2) for the Newtonian case. In the opposite limit when the viscoelastic relaxation time is large (or $\omega_r$ is small), $\omega_r \rightarrow 0$, the second term on the left hand side of the Maxwell model becomes singular implying that the model has to be augmented by the inertial terms. Furthermore, even if the inertial terms are insignificant, this singularity contributes to the rod dynamics at the short time scale [103]. An analysis of these effects requires modifications of the MRS experimental protocol.

In the Kelvin-Voigt model, the second term on the left hand side corresponds to the elastic torque. Since the spring and dash pot are connected in parallel, both elements have equal deformations. Therefore, the Kelvin-Voigt model reduces to the Newtonian case, i.e. eq.(2), when the viscoelastic relaxation time is large (or $\omega_r$ is small), $\omega_r \rightarrow 0$. Again, contrary to the Maxwell model, as the viscoelastic relaxation time decreases (or $\omega_r$ increases), $\omega_r = G/\eta \rightarrow \infty$, the elastic resistance of the material to the rod rotation becomes much stronger than the viscous resistance. This results in a singularity implying that the Kelvin-Voigt model also has to be augmented by inertial terms and some special care must be taken to analyze the rod dynamics in this case.

### 5.6 STATIC MAGNETIC FIELD

We first consider the simplest case of the rod orientation by a static magnetic field applied in the y-direction assuming that the rod magnetic moment initially points in the x-direction. Substituting the relation $\theta + \varphi = \pi/2$ into eqs.(13)-(14), one can obtain the time evolution of the rod orientation for the Newtonian, Maxwell and Kelvin-Voigt fluids, Fig. 5.

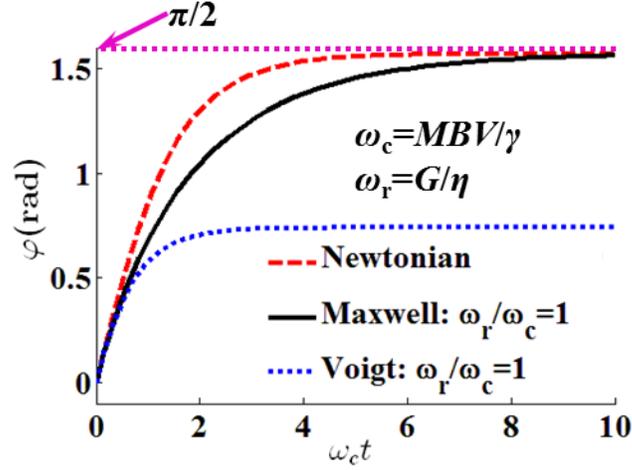

Figure 5. Time evolution of the rod orientation for Newtonian, Maxwell and Kelvin-Voigt fluid.

The behavior of the rod rotation in the Maxwell fluid appears to be similar to that in the Newtonian fluid. In the Maxwell fluid, it requires more time for the rod to reach the equilibrium than one could expect for a Newtonian fluid with the same viscosity. Remarkably, in the Kelvin-Voigt fluid the rod does not co-align with the applied field: the Kelvin-Voigt spring pulls the rod back to its initial position hence the rod makes angle $\varphi_{eq}$ which can be found by letting d$\varphi$/d$t$=0 in eq.(14) and applying the relation $\theta + \varphi = \pi/2$:

$$\omega_r \varphi_{eq} = \omega_c \cos \varphi_{eq} \tag{15}$$

Therefore, one can distinguish a Kelvin-Voigt fluid from a Newtonian or Maxwell fluid by applying a static magnetic field perpendicular to the initial orientation of the rod and checking the equilibrium orientation of the rod. To distinguish between Newtonian and Maxwell fluid, one needs to analyze the dynamics of the rod rotation and compare it with the predictions of these models.

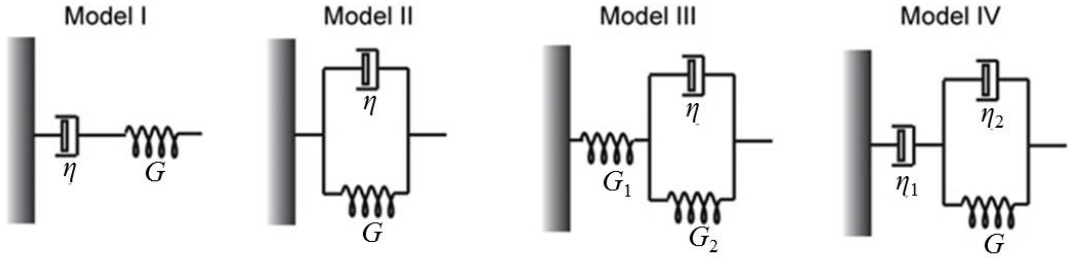

*Figure 6. Models used in Ref. [65] to characterize the rheological properties of nucleus in the living cell.*

In some studies, more complicated models were applied[65]. Figure 6 is taken from Ref.[65] to illustrate different models applied to interpret the viscoelastic properties of a nucleus of a living cell. In Ref.[65], the nucleus was found to be neither Maxwell nor Kelvin-Voigt type fluid. The time evolution of the orientation of the nanorod was analyzed and model IV was proven to be the best fit for the nucleus. In Ref.[101], the same model IV was found to be also suitable for description of the rheological properties of the cell cytoplasm.

## 5.7 ROTATING MAGNETIC FIELD

There is a very limited amount of works that use rotating magnetic field to characterize the materials' viscoelasticity[71]. Consider the specifics of the particle rotation when the applied magnetic field revolves at angular frequency $\omega$. In this case, the relation $\theta + \varphi = \omega t$ holds true. Substituting this relation into eqs.(13)-(14), the evolution of the rod orientation under a rotating field can be obtained.

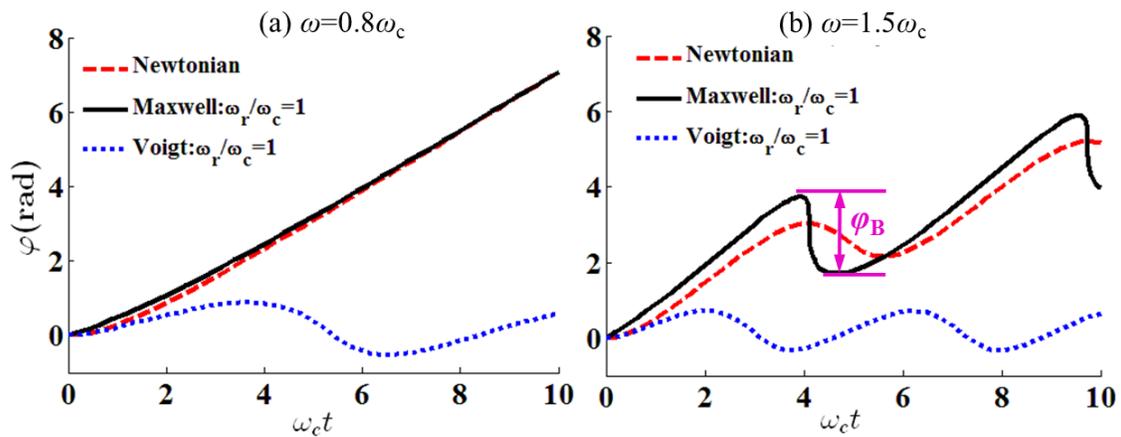

*Figure 7. Time evolution of the rod orientation for the Newtonian, Maxwell and Kelvin-Voigt fluids under a rotating magnetic field. (a) $\omega<\omega_c$, (b) $\omega>\omega_c$.*

Figure 7 shows time evolution of orientation of a ferromagnetic rod in three different fluids under two different rotating frequencies. A transition from synchronous to asynchronous rotation shows up in a Maxwell and Newtonian fluids. While the two transitions are similar, the rod trajectory in a Maxwell fluid demonstrates a slight skew relative to the rod trajectory in a Newtonian fluid. The transition occurs at the same frequency $\omega_c$ for both fluids. Therefore, the critical frequency $\omega_c = 2\pi f_c$ cannot be used for characterization of fluid elasticity. Moreover, as shown in Ref.[71], the average frequency of nanorod rotation (averaged in asynchronous regime over the period of the nanorod swinging back and force) in the Maxwell and Newtonian fluid is the same! The only distinguishable difference between rod rotations in the Maxwell and Newtonian fluids in the asynchronous regime is that the backward rotation ($d\varphi/dt<0$) in the Maxwell fluid is faster due to the additional restoring force from the spring.

Due to the presence of a parallel spring in the Kelvin-Voigt fluid, the rod will always oscillate during rotation. Therefore, it is easy to distinguish the Kelvin-Voigt fluid from the other two. The phase diagram, Fig. 8, is taken from Ref.[71]. It shows different rotation regimes of a paramagnetic rod in a Maxwell fluid. Although experiments were performed with paramagnetic nanorods, similar conclusions can be drawn for a ferromagnetic rod. The ($\omega_c$, $\omega$) - plane corresponds to a Newtonian fluid and the ($\omega_c/\omega_r$, $\omega$)- plane corresponds to a pure elastic material. Figs. 8 (a) and (b) illustrate the transition between the two regimes in Newtonian fluids. Figs. 8 (c) and (d) correspond to the oscillations in the vicinity of the equilibrium position for an elastic material under a rotating field. Figs. 8 (e) and (f) depict the asynchronous rotation of a rod in a Maxwell fluid showing a faster backward rotation compared to a Newtonian fluid.

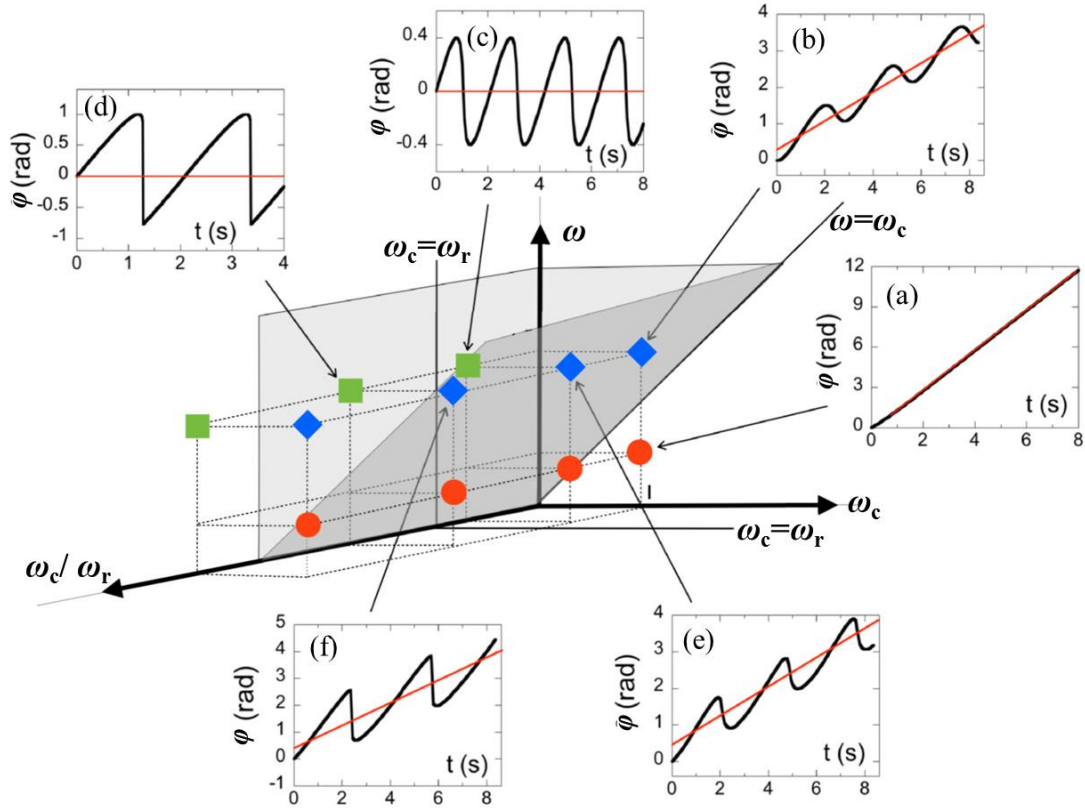

*Figure 8. Phase diagram[71] showing different rotation behavior of the nanorod. The black line is the trajectory of the nanorod and the red line shows the average angular velocity.*

Ref.[71] reports a detailed analysis of rotation of a paramagnetic rod in a surfactant wormlike micellar solution cetylpyridinium chloride (CP+; Cl−) and sodium salicylate (Na+; Sal−) (abbreviated as CPCl/NaSal) dispersed in a 0.5MNaCl brine. This solution follows the Maxwell rheological model. The following major difference between the Maxwell and Newtonian fluids is revealed: the amplitude of the backward rotation $\varphi_B$ in a Maxwell fluid (marked in Fig. 8(b)) should reach a plateau while the frequency $\omega$ keeps increasing. In contrast, the angle $\varphi_B$ for a Newtonian fluid keeps decreasing and finally reaches zero (see Fig. 9).

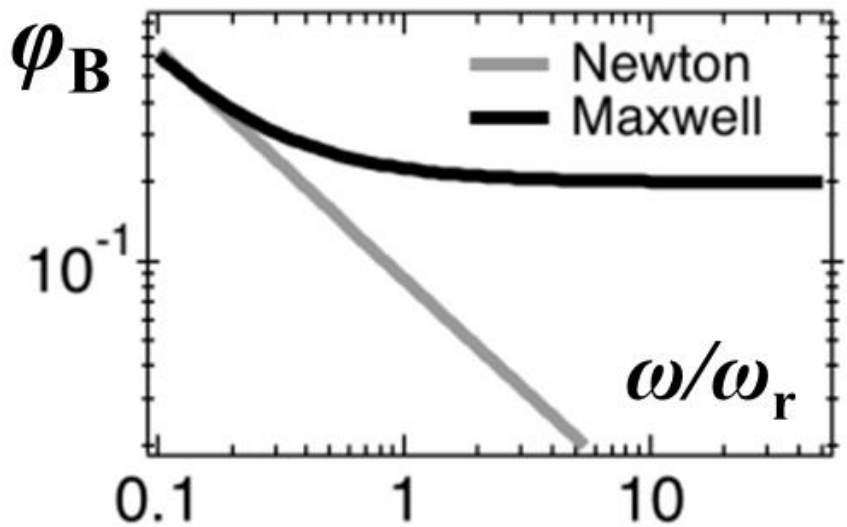

*Figure 9. The frequency dependence of $\varphi_B$ for the Newtonian and Maxwell fluids.(Taken from ref.[71])*

The MRS experimental results from Ref.[71] show excellent agreement with the measurements obtained from a cone-and-plate rotational rheometer. As such, MRS appears to be a promising technique for characterization of viscoelasticity of these fluids.

# 6 EXPERIMENTAL AND INSTRUMENTAL METHODOLOGY

## 6.1 MAGNETIC ROTATIONAL SPECTROMETER

An experimental realization of MRS can be done with a microscope and a high speed camera. A transparent cuvette containing the studied material with the dispersed magnetic nanorods (nanoprobes) is placed on the stage of an up-right optical microscope. The microscope is coupled with a high speed camera connected to and controlled by a computer. A Motorized XY translation stage adjusts the position of the microscope stage whereas a linear piezoelectric stage controls the height of the cuvette with the sample. Some groups employ permanent magnets fixed on a moving stage that are spun to produce the AC field[65, 104]. Other groups utilize current carrying coils that generate magnetic fields with a broad frequency band[67-69, 71, 74, 76, 83, 85, 105-108].

The permanent magnets are mostly used when one needs to generate high gradients of the magnetic field[65, 104, 109-116]. When the stage with a magnet rotates or moves back and forth, the nanoparticles readily follow the generated field. This method, however, makes it difficult to estimate the strength of the magnetic field at the point of interest and to accurately rotate the magnetic field with high frequency. Typically, the frequency band generated by the moving stage is very narrow limiting the application of permanent magnets.

In applications that require an AC field covering a broad frequency band, electromagnets are more attractive. The magnitude, direction, and frequency of the magnetic field can be remotely controlled by simply changing the current on the magnetic coils [51, 64, 67-69, 71, 74, 85, 94, 106, 114][83, 101, 117-120]. Several groups have been working in the past years on the development of electro-magnetically controlled optofluidic devices[51, 64, 67-69, 71, 74, 85, 94, 106, 114][83, 101, 117-124]. It has been demonstrated that the generation of the rotating magnetic field can be done with small Helmholtz coils. These coils provide the rotation frequency in the range between 1 Hz to 1 kHz and the field strength in the range of tens of milli-Tesla. In applications that require stronger magnetic fields, the generation of the rotating magnetic field can be done with the coils containing ferrite cores[67-69, 83, 87].

So far, different groups attempted different methods of creating magnetic field. One can employ one or two electromagnets in the Hull configuration with the square wave-like changes in the magnetic field[61, 70, 106], two perpendicular - or four connected in pairs - electromagnets with the square wave-like changes in the magnetic field[76], and two perpendicular – or four connected in pairs - electromagnets with 90 degree shifted sine oscillations of the magnetic field [68, 71, 125-129].

Magnetic generation of the square wave-like field is convenient in a laboratory setting, but its frequency should be set with a precaution to accommodate for various experimental features. For instance, the temporal resolution of a MRS measurement needs to be high in order to measure the characteristic time constant of the probe rotation. For example, with a high speed camera providing 180 frames per second in the bright field, the characteristic time scale of rotation would have to be larger than 50 milliseconds to capture at least 10 frames per transition. The frequency of the generator would have to be low enough to fit in that timeframe. This estimate is just one example illustrating the reasoning behind setting the wavelength of the square wave generator.

To the best of our knowledge, no attention had been given to the effects of the z-component of magnetic field on MRS. In a lab setting, in addition to the Earth's z-component magnetic field, there are ambient magnetic fields generated by the lab equipment that need to be taken into account or canceled. In particular, a magnetic bar generating a rotating magnetic field will always create a z-component of magnetic field. Therefore, it seems inappropriate to use this method of generation of the rotating field for the accurate MRS analysis. A precisely controlled and measured rotating field with an active cancelation of the z-component of the field seems to be the only way that would yield reliable MRS results and be compatible with a large variety of probes and liquids.

Our group employs four coils that are arranged face-to-face and are used to create a rotating magnetic field in the focal plane of the cuvette. A fifth coil is oriented vertically (coaxial with the optical axis of the microscope) and is positioned directly under the cuvette to control the Z-component of magnetic fields. The sample can be illuminated either from the bottom (for transmitted light microscopy) or from the top (for dark field microscopy) (For illustration, see Fig. 10).

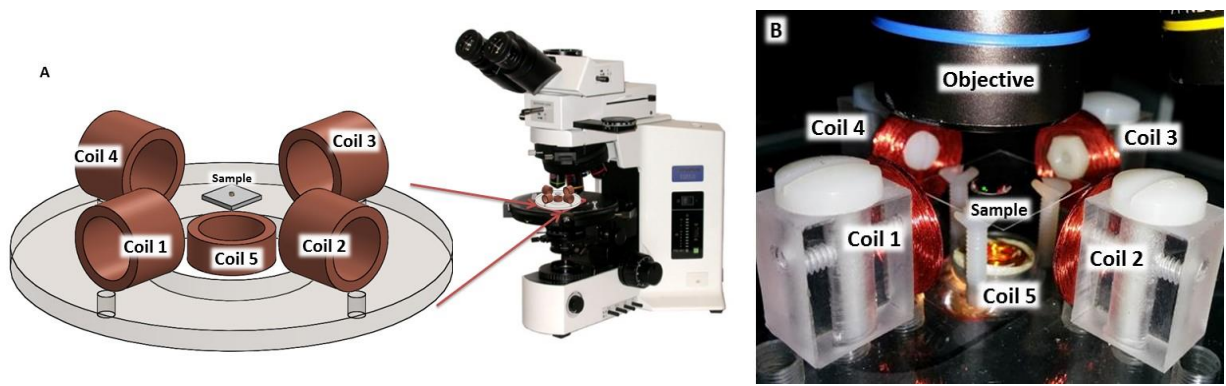

*Figure 10. The MRS apparatus, as realized in our group; (a) is the schematic and (b) is the actual set-up. Coils 1, 2, 3, and 4 are used to cancel out the ambient magnetic field in the X and Y directions and rotate the nanorods. Coil 5 is to cancel the ambient magnetic field in the Z direction. The sample rests on 3 supports, allowing to control the tilt of the sample. The apparatus is placed under the objective of a microscope, thus allowing for direct observation of rods' orientation.*

The coils are able to generate different spatial and temporal patterns of the magnetic field in micro- and milli-Tesla range. This level of control is achieved with a combination of precise and flexible current generation and real-time three-dimensional magnetic field measurement. In order to accurately and flexibly generate a current for the coils, we employ a programmable voltage generator coupled with a voltage-to-current converter for each coil. We are thus able to generate any waveform of current through any of the 5 coils.

The system, however, requires further calibration. Due to coils' and amplifiers' imperfections, the same signal sent to two different coils can produce magnetic fields that sometimes differ by a factor of two. Moreover, due to geometrical imperfections of the set-up, coils that are oriented along the x-axis of the stage produce the y- and z- components of the magnetic field. Finally, ambient magnetic fields in a lab that uses magnets or electronic equipment can be very strong, and when coupled with the magnetic field of the Earth, can invalidate the experiment. It is thus impossible to know without proper calibration what field is produced in the sample. To combat all of the aforementioned obstacles, we employ a three dimensional magnetic sensor located in place of the sample during calibration. The current passing through each individual coil is consecutively varied and the magnetic field in each axis is measured. We then linearly fit the data and use the obtained information to calculate what signal to send to the coils to create a uniformly rotating magnetic field. Fig. 11 illustrates the importance of such a calibration; it can be seen that in order to create a uniformly rotating magnetic field (a), the signal generator must generate a non-trivial signal for each of the five magnets (b).

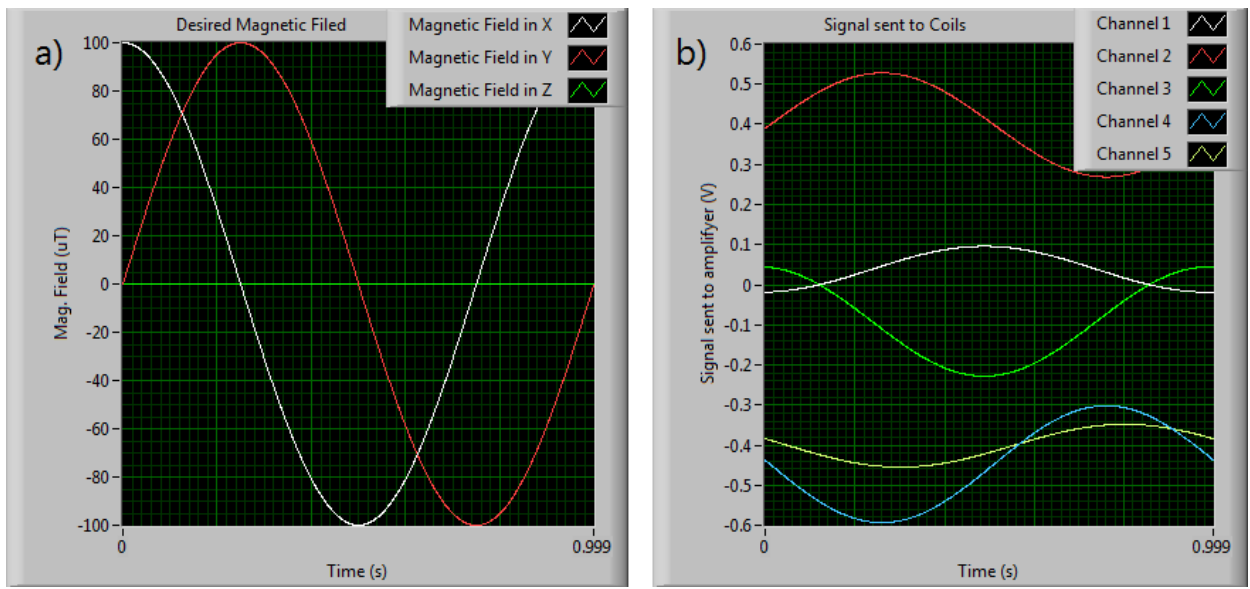

*Figure 11. Results of calibration of magnetic field. (a) Desired uniform magnetic field. (b) Signal that the signal generator must produce in order to achieve the uniform magnetic field specified in (a).*

In MRS applications, one relies on the optical imaging of the rotating magnetic probes. Therefore, the system has to meet the following criteria: 1) the size of the stage with integrated magnetic coils has to be small enough to place it under the microscope objective; 2) the setup must be flexible to ensure an easy rearrangement of the coils or alternation of the distance between them and should be able to eliminate any external magnetic perturbation during measurements; 3) the spatial distribution of the generated magnetic field has to be known to ensure a proper correlation of the field with the probe motion.

Upon induction of the two-dimensionally rotating magnetic field, the nanorods rotate in the field plane following the field vector. By varying the frequency of rotation of the applied magnetic field, one can probe different mechanical reactions of the sample by filming the nanorod motion and interpreting the nanorod behavior with a set of rheological models.

Materials and magnetic probes are studied in the bright and dark field modes. Polarization and fluorescence microscopy are also used to examine different features of the phenomenon in question. For example, a domain structure of evaporating aqueous solution of polyethylene oxide (PEO) can be imagined using polarization microscopy. In Fig. 12, we illustrate this structure with the embedded nickel nanorods. MRS imaging determined that the viscosity of the solution increases dramatically before crystallization occurs. This thickening effect is advantageous for MRS because it allows the observer to follow the nanorod rotation not changing the focal plane: the nanorod stays in focus all the time.

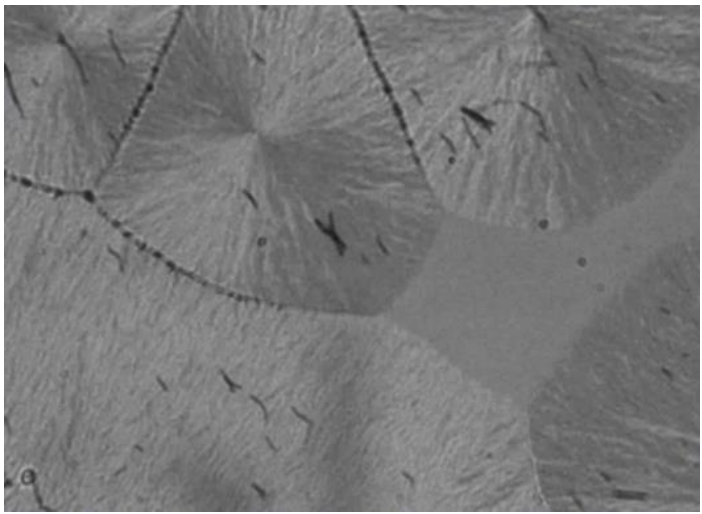

*Figure 12. Growing polyethylene oxide crystals through nickel nanorods, imaged under polarized microscope. The crystal boundary, as it grows, does not move the nanorods. Growth of ice in water, conversely, is known to move objects.*

Typically, one selects a spot containing several nanorods in the focal plane and then follows their rotation using some tracking algorithm [94]. Having many nanorods in the same video, one can collect statistics and confirm the correctness of the collected data. The method of MRS with magnetic nanorods does not employ any mechanical actuators. The instrument features high sensitivity and millisecond response times to changes in environment or external magnetic field.

## 6.2 PROBES

In general, different MRS techniques are developed with specific types of magnetic probes in mind. The most common probes are either spherical[62, 130, 131] or elongated probes[71, 125, 132].. In this section, we will briefly go over different types of probes, their strengths and weaknesses.

### 6.2.1 Spherical probes

Many groups use commercially available magnetic beads. These beads can be made half-covered with a florescent paint on one side. By rotating the probe with a magnetic field, either the fluorescent side or the non-florescent side is facing the camera. Since early 2000s when Kopelman's group developed the MagMOONs (magnetically-modulated optical nanoprobes), these spherical probes become quite popular. The bead diameter is measured in a few microns with one side coated with either vapor- deposited aluminum or sputter-coated gold[62, 63, 127]. One can track the 'blinking' of the particles as the fluorescent side of the spheres flips back and forth. Detection of the probe rotation requires high quality optics to differentiate the wanted fluorescence from unwanted reflections and scattered light. The method relies on the ability to subtract the background image from a fluorescent particle image and if the background is changing rapidly due to an ongoing reaction in the sample, it is extremely difficult to do.

Important shortcoming of multidomain spherical probes is the fact that the magnetic moment has no preferred geometrical orientation and is free to rotate within the probe, introducing a viscosity-dependent error into the measurement. Overall, while this method provides only a rough measurement of viscosity, it allows one to monitor with a very high resolution the spatial variation of viscosity.

Thus, this method appears attractive to assess drastic changes of viscosity at particular locations. For example, asynchronous rotation of spherical beads was used to accurately determine bacterial growth within a microchannel[128]. The setup consists of two very small coils described in Ref.[106]. The apparatus does not seem to measure the viscosity very accurately, but does detect large changes in viscosity[106].

### 6.2.2 Elongated probes

There are several approaches to make magnetic probes elongated. Among them are the field directed assembly of magnetic nanobeads or electrostatic complexion between oppositely charged particles[37, 114, 133-136], method of filling and decorating nanotubes with magnetic nanoparticles [85, 137-139], template based electrochemical deposition[38, 140-142] as well as template free wet chemical synthesis[41, 42, 143]. To make fluorescent elongated probes, one deposits plastic beads with embedded magnetic nanorods onto one glass slide and fluorescent dye onto another glass slide, and consequently mechanically rubs the two slides together[132]. Thus, the dye mixes with the beads and the beads become elongated due to deformation of the plastic forming rolls. These rolls may join together to form longer rolls. The method is simple to perform and requires only commercially available ingredients. However, one should use it with a precaution because the measurements with these probes are highly uncertain: it is impossible to control magnetic or geometric properties of the resultant probes[101].

Bacri's lab developed a method to produce chains of magnetic nanoparticles for the purpose of intracellular rheological measurements[70]. The elegance of the method roots in the fact that the chains self-assemble in the living cells due to the cells' digestive systems. The probe formation process is as follows. First, the cells are fed with magnetic nanoparticles of 50-100nm, which enter the cells via endocytosis. As a result of this process, the nanoparticles end up covered with endocytotic membrane and due to internal flows in the cell, the nanoparticles flow into larger organelles, namely, endosomes (600 nm in diameter). Filled with magnetic nanoparticles, the endosomes become magnetic and self-assemble into magnetic chains. These chains then served as probes, which were rotated by an external magnetic field to measure viscosity inside the cells. These probes allowed to conduct very unique experiments [100, 101]. However, it is very difficult to specify geometrical and magnetic properties of the chains. In order to characterize these chains, the group had to mechanically lyse the cells to extract magnetic particles and analyze the chain properties using known media. Since the size of endosomes had a broad distribution, the calibration was very approximate and provided only rough estimates of the rheological properties of the cellular biofluid.

The template based electrochemical growth of nanorods from magnetic metals and method of filling nanotubes with magnetic nanoparticles appeared most attractive [140]. In this method, nanoporous alumina is used as a template and metal nanorods are electrochemically grown inside pores of this membrane; or carbon nanotubes are grown inside these pores and subsequently filled with a magnetic fluid and then the carrier fluid was evaporated[85]. The electrochemical growth of magnetic nanorods enables one to precisely control the size of the nanorods [38, 67, 69, 94]. One can generate Ni, Co, permalloy, and other metallic nanorods that can be used for MRS. As an example, the SEM image shown in Fig. 13 shows cobalt nanorods produced in our laboratory. These nanorods and the magnetic-fluid filled carbon nanotubes [85, 137] can be produced in the uniform sizes and manipulated via an external magnetic field. Magnetic nanorods of 100 nm - 200 nm diameter can be covered with polymers to prevent their agglomeration[94, 116, 142]. The template grown nanorods exhibit ferromagnetic order, with a distinguishable hysteresis loop (Fig 14). The nanorods have a very well defined radius that is set by the template and a narrow length distribution, which allows for simple and precise calibration in a known liquid. The probes of the microscopic size can be easily produced by the electropolishing technique[144, 145].

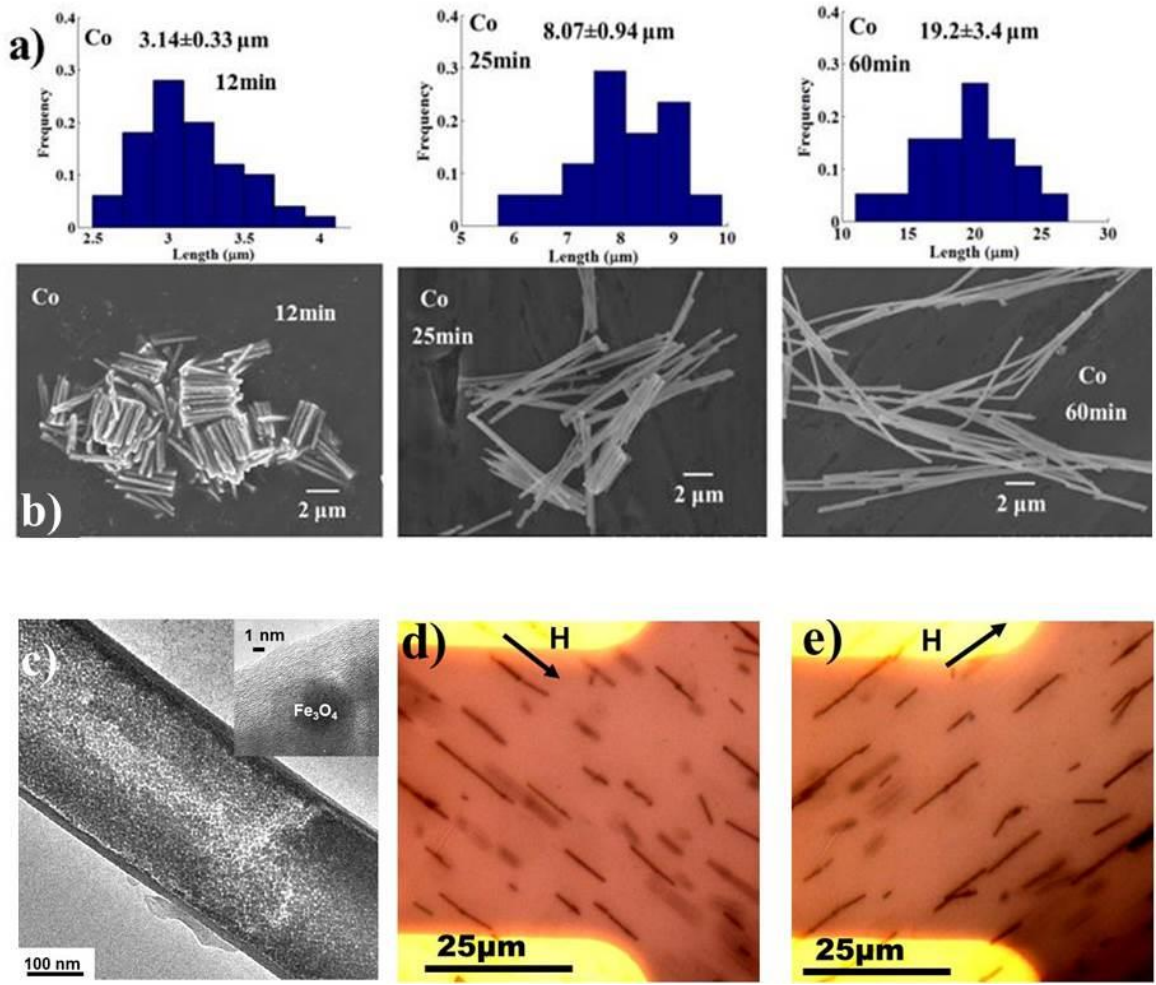

Figure 13 (a) The length distribution and (b) SEM micrographs of electrochemically grown cobalt nanorods produced at different deposition times; (c) TEM image of carbon nanotube filled with dispersion of $Fe_3O_4$ nanoparticles; (d)-(e) ordering of magnetic nanotubes in the plane of a wafer [85].

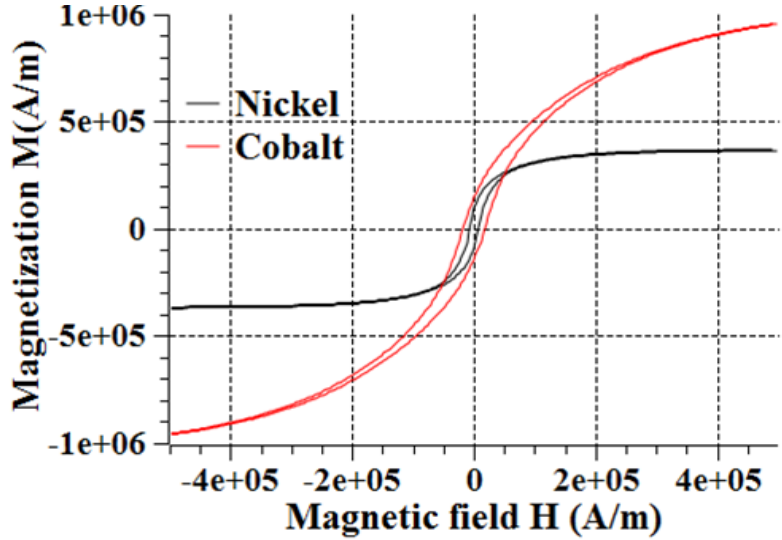

Figure 13. Hysteresis loops for nickel and cobalt nanorods.

### 6.2.2.1 Selecting the right probes for particular applications

A set of magnetic coils in MRS are usually designed to cover a special range of magnetic fields. Therefore, liquids with different viscosities require different probes. For MRS applications, the dependence of the dimensionless frequency of rotating magnetic field $V = 2\pi\gamma f / (mB) = 8\pi\eta f (l/d)^2 / [3MB(\ln(l/d) - A)]$ on the materials properties of the probe is of particular importance. An analysis of this dependence allows one to choose an appropriate material to study the liquid in question at the given field strength. Table 1 provides the values of saturation magnetization M for the most popular materials and their Curie temperatures and Table 2 lists viscosities of common liquids illustrating the MRS challenge.

We consider the effect of weak and strong magnetic fields taking B=0.0015T as an example of the weak fields, and B= 0.01T as an example of the strong fields. Fgure 15 illustrates the dependence of V-parameter on fluid viscosity. The dashed line separates the region of synchronous rotation (V<1) from the region of asynchronous rotation (V>1) of magnetic nanorods. According to the graphs, to probe liquids with viscosities less than ~ 600 mPa·s, one could use weak magnetic fields and pick nickel or ferrite nanorods from a chosen series of materials. Fluids with greater viscosities, up to 1500 mPa·s, can still be probed by the weak fields, but one needs to apply stronger magnets or use more magnetic probes, such as iron and cobalt nanorods.

Fluids with even greater viscosities up to 10000 mPa·s require strong magnetic fields which decrease V-parameter. When the field is increased to B = 0.01T, the high aspect ratio nanorods (e.g. $l/d$=15) made of any materials listed in Table 1 should be able to probe these fluids.

Table 1. Saturation magnetization M (room temperature) and Curie temperature of the most popular magnetic materials used for nanorods synthesis. (Compiled from Refs.[82, 146])

| Substance | Magnetization M (K A/m) | Curie temperature (°K) |
| --- | --- | --- |
| Iron (Fe) | 1707 | 1043 |
| Cobalt(Co) | 1400 | 1400 |
| Nickel (Ni) | 485 | 627 |
| Magnetite ($Fe_3O_4$) | 480 | 858 |

Table 2 Viscosities of some liquids of interest.

| Liquid | Viscosity ( m Pa·s) |
| --- | --- |
| Ethanol | 1.74 |
| Ethylene glycol | 16 |
| Glycerol | 1200 |
| Molten glass | > 10000 |
| Water | 1 |

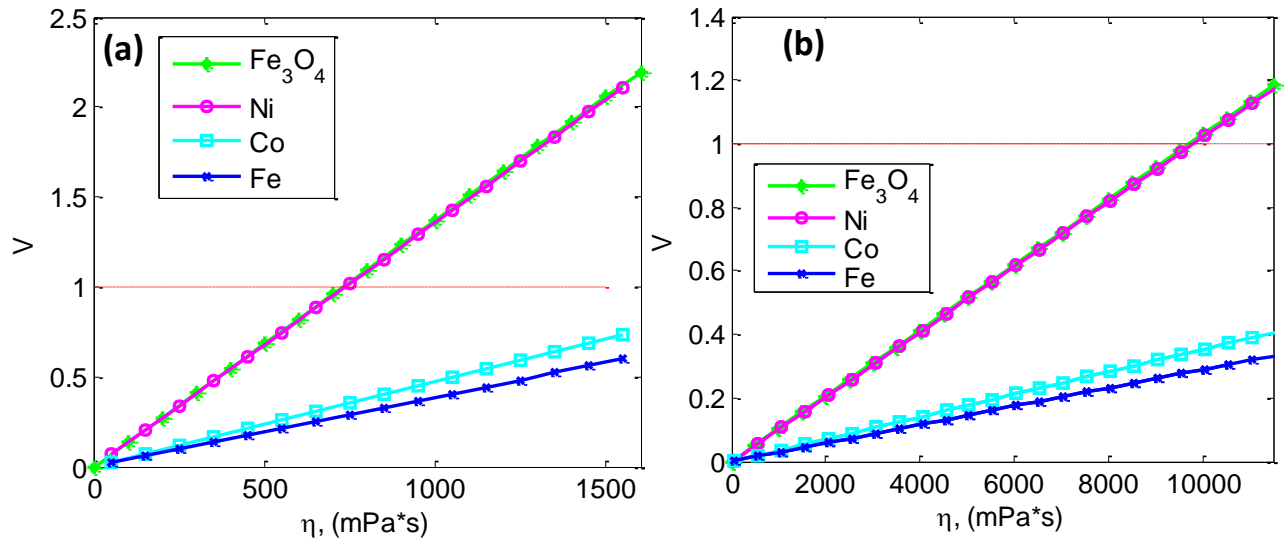

*Figure 15. The dependence of V-parameter on the viscosity of carrying fluid for nanorods with aspect ratio l/d=15 (a) magnetic field B=0.0015T (b) magnetic field B = 0.01T*

### 6.2.2.2 Characterization of rod-like magnetic probes using MRS: self-calibration

The rod-like probes with a high aspect ratio are much more attractive for MRS applications because of their easy manipulation and detection. As follows from eq. (6), the critical frequency depends only on the aspect ratio of the probe. This ratio can be made high to diminish the error in the length and diameter measurements. For example, making a nanorod with the diameter of d = 200nm and length l = 20 μm one can easy obtain d/l = 100. At this high aspect ratio, small errors in optical measurements of the nanorod diameter and length would not influence determination of the critical frequency given by the right hand side of eq. (6).

From eq. (6) it also follows that MRS experiments allow for direct characterization of magnetic properties of nanorods that are used for the rheological analysis. Measuring the aspect ratio of the probe and the field distribution in the focal plane of rotating nanorod, one can determine magnetization of the probe prior to measuring the rheological properties of the material [67-69, 83, 85, 94]. The procedure is as follows.

Nanorods are dispersed in a volatile liquid of known viscosity (for example, ethanol or water). Then one runs a MRS experiment to measure $f_c$ for different nanorods. Solving eq.(6) for M, and using the measured parameters, one can determine the average magnetization of the nanorods. Since one uses an assembly of nanorods with a narrow *l/d* distribution, the average M is obtained accurately. After measurements, the humidity in the environmental chamber is decreased to let the liquid evaporate. The nanorods are typically adhered to the substrate by a weak van der Waals force. To prevent this adhesion, a linear piezoelectric actuator under the substrate oscillates at ultrasound frequency, thus perturbing the drying nanorods. Then, after filling the cuvette with the material in question, the sample is ready for characterization. The validity of the results on nanorod magnetization obtained using the self-calibration procedure was examined against the results on nanorod magnetization obtained with the Alternating Gradient Magnetometer (Princeton Measurements Corp) [68, 85, 94]. Both sets of results show good agreement thus allowing one to confidently apply the self-calibration procedure.

# 7 KEY RESEARCH FINDINGS

## 7.1 CHARACTERIZATION OF FLUIDS WITH LOW VISCOSITY

As follows from eq. (6), the measurements of viscosity of thick fluids can be done at low frequency of the rotating magnetic field. Thin fluids, meanwhile, require application of a high frequency. The rheological analysis of thin fluids presents a challenge even on macroscopic samples[34, 95]. One can address this problem with a special care by choosing the coils that provide milliTesla fields and by selecting a suitable camera. As an illustration of the robustness of MRS and its ability to deal with low viscous fluids, Fig. 16, presents the results of measurements of viscosity of butterfly saliva and a set of MRS viscosity data on aqueous solutions of sucrose[68]. Prior to butterfly saliva measurements, the MRS tool was calibrated on 10%-40% sucrose solutions and yielded viscosity measurements that agreed with published data for sucrose solutions[147]. In insects, saliva lubricates the mouthparts, aids digestion, and dissolves viscous and dried substances[148, 149]. We did not observe any viscoelastic effects or non-Newtonian behavior of saliva droplets. Unlike mammalian saliva containing high molecular weight mucins [150], butterfly saliva has much simpler chemical composition which does not include mucins[148].

These results suggest that saliva should not be needed for liquefying nectars with sugar concentrations up to 30%-40%; viscosity stratification would not be expected when butterflies feed on nectar with 30-40% sugar concentrations. With MRS, we were able to draw the very important biological conclusions that changed the textbook knowledge and posed new questions on the mechanisms of uptake of liquid food by Lepidoptera[151].

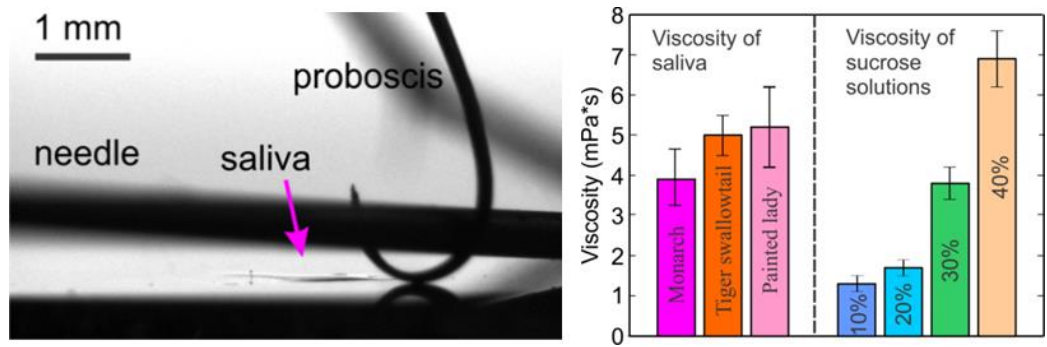

*Figure 16. Obtaining nanoliter droplet of butterfly saliva. Black bar crossing the image is the needle used to extend the proboscis of a live butterfly. The table presents the comparison of the viscosity of butterfly saliva with that of sucrose solutions of different concentrations[68].*

## 7.2 CHARACTERIZATION OF THIN FILMS THICKENING WITH TIME

In many cases when polymers crosslink to form a gel, viscosity changes exponentially fast[97]. In biological and biomedical applications, the characterization of films thinner than 100 μm is a challenge. MRS allows one to make a step forward and study polymerization in such thin films. As an illustration of the robustness of the MRS technique, we investigated the time-dependent rheology of microdroplets of 2-hydroxyethyl-methacrylate (HEMA)/diethylene glycol dimethacylate (DEGDMA)-based hydrogel during photopolymerization synthesis [69, 96], Fig. 17.

Employing optical spectroscopy, one can study the mechanisms of viscosity change. For example, HEMA polymerizes through the carbon-carbon double bonds and crosslinks through the two double bonds in DEGDMA. Therefore, following the rate of decrease of the carbon-carbon double bonds in the system by measuring the rate of disappearance of the 1635 cm$^{-1}$ peak corresponding to the carbon-carbon double bonds, one can monitor the crosslinking and correlate it with the rheological data[69].

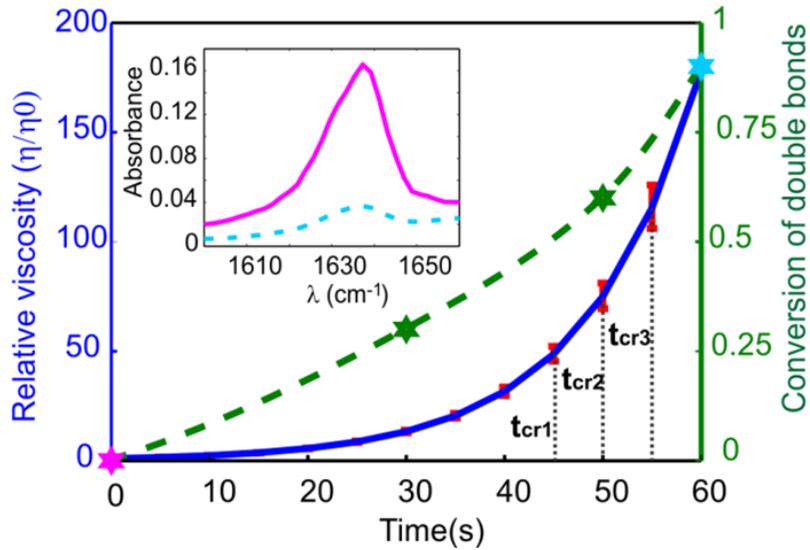

*Figure 14. The solid line illustrates the change of relative viscosity ($\eta_0$ is the solution viscosity prior to polymerization) with time of photopolymerization of the 4.5 wt% crosslinker solution (the left y-axis) and the stars show the conversion of the double bonds during the photopolymerization (the right y-axis) and the dashed line is a trend-line. The inset shows a FTIR spectrum of the solution near 1635 cm-1 before (the solid line) and after 60 seconds of polymerization (the dashed line)[69].*

In many applications, the material thickens upon evaporation of the solvent[5] . The MRS technique is also applicable to study rheological properties of thin films and droplets. In Fig. 18, we present the analysis of thickening of the aqueous solution of mullite ($3Al_2O_3 \cdot 2SiO_2$)[152] [73]. Applying the MRS analysis at different time moments during drop evaporation, one can infer an exponential dependence of viscosity on time. The phenomenological parameters of this complex liquid were measured by fitting the experimental data points using this exponential approximation. Moreover, employing a model of drop evaporation[152], one can relate the change of viscosity with the mullite concentration. This information is very important for interpretation of the thickening mechanism in materials with a complex structural organization, where the gelation mechanism involves multiple bonds.

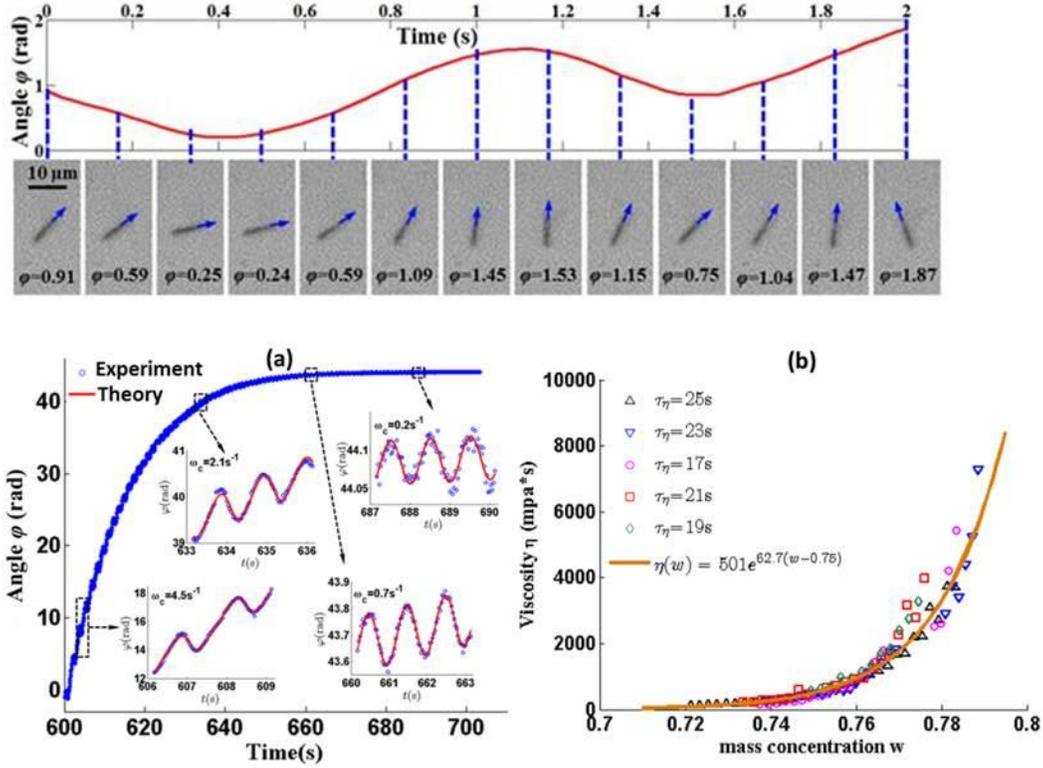

*Figure 15. A gallery of images showing oscillations of the nanorod inside evaporating mullite sol. a). The time evolution of the magnetization vector spinning inside evaporating mullite droplet. The angular frequency of the magnetic field is ω = 2π s$^{-1}$. The blue circles are the experimental data points extracted from the video and the red lines are the theoretical curves. b). Viscosity of the mullite solution as a function of mass concentration of mullite [73].*

## 7.3 Measurements of Interfacial Viscosity

Interfacial interactions play a dominant role in biological systems ranging in size from cellular structures to small insects and are of great interest to biologists and engineers alike. [153, 154] Traditionally, interfacial viscosity measurements employed macroscopic probes[155-158], which are not very sensitive as they strongly interact with the bulk material. The ratio between the interfacial and bulk viscous drags is characterized by the Boussinesq number (Bo)[153]:

$$Bo = \frac{\text{interfacial drag}}{\text{bulk drag}} = \frac{\eta_s P / L'_c}{\eta A / L''_c} \tag{16}$$

where $\eta$ is the bulk viscosity, $\eta_s$ is the interfacial viscosity, $P$ is the probe's contact perimeter with the interface, $A$ is its contact area with the bulk, and $L'_c$ and $L''_c$ are the characteristic lengths over which the interactions take place in the interface and bulk, respectively. It follows from this equation that as the ratio $(P/L'_c)/(A/L_c")$ becomes large, the measurement becomes more sensitive to the interfacial drag. Advancements in MRS have allowed using micron-scale probes, greatly increasing the Boussinesq number and making the measurements more sensitive[37]. This allowed to visualize deformations of fluid interfaces under applied stress and correlate the structure and rheology in monolayer films [159, 160].

# 8 CONCLUSIONS

In-plane rotation of magnetic particles (probes) in a rotating magnetic field has a characteristic feature: as the rotation frequency of the applied field increases, the particles first rotate in unison with the field (synchronous rotation) and then when the frequency of rotation of the external field passes some characteristic frequency, the particles undergo a transition from synchronous to asynchronous rotation. This transition depends on the fluid viscosity. Therefore, one can take advantage of this effect and employ it for characterization of the viscous properties of different materials. In this paper, we review the theory of transition from synchronous to asynchronous rotation of particles and discuss its experimental implementation. This effect laid the ground for development of a new method which we call Magnetic Rotational Spectroscopy. In MRS, one studies the rotation of magnetic particles by scanning over the frequency of the applied rotating magnetic field. MRS can be used for the *in-situ* (or *in-vivo*) rheological measurements of complex fluids. Spherical probes allow for testing of very large changes in viscoelasticity, while elongated probes or rod-like probes allow for very accurate measurements of small and medium changes of viscosity and elasticity separately. The method employing the rod-like probes relies on imaging of large rotations of magnetic rods, making its tracking algorithms stable and able to provide reliable data. Moreover, MRS with long probes allows for characterization of not only of Newtonian viscous fluids, but also of viscoelastic Maxwell or Kelvin-Voigt fluids. Only small quantity of the sample is needed: for example, taking that the rotating 5 μm long nanorod can cover the 1pL volume, one can measure the rheological property of a drop of comparable size. The material rheology can be probed on micro as well as on nano levels, depending on the size of the used nanorods. Remarkably, an increase of viscosity can be traced beyond the point when the material undergoes transition to a gel and the domains start to appear. In conjunction with carefully controlled the rotating magnetic field to prevent unwanted bias and oscillations, MRS with rod-like probes provides unprecedented control over micro and nanoscale rheological measurements. MRS is thus an irreplaceable tool not only in characterization of newly synthesized materials, scarce natural materials, and thin films, but also in understanding of their internal dynamic processes. We expect that this method will open new horizons in the quantitative rheological analysis of fluids inside the living cells, microorganisms, and aerosol droplets with thickeners.

In summary, we believe that MRS is a robust, flexible, and accurate method capable of characterization of nanoliter samples of materials with complex and/or time-dependent rheology. This method is sufficiently simple to implement with inexpensive microscopes and magnetic coils, yet sufficiently versatile to accommodate a broad variety of experiments.

# 9 ACKNOWLEDGMENTS AND REFERENCES

## 9.1 ACKNOWLEDGMENTS

This work has been conducted over the last decade and we thank our collaborators, especially, Guzelia Korneva, Derek Halverson, Gary Friedman, Yury Gogotsi, Alexey Aprelev, Taras Andrukh, Daria Monaenkova, Binyamin Rubin, Igor Luzinov, Bogdan Zdyrko, Ruslan Byrtovyy, Jeffery Owens, Kim Ivey, and David White. The authors are grateful for the financial support of the National Science Foundation and the Air Force Office of Scientific Research.